\documentclass[floats,floatfix,showpacs,amssymb,prd,twocolumn,superscriptaddress,nofootinbib,nolongbibliography,reprint]{revtex4-2}

\usepackage{amssymb,amsmath,verbatim,mathtools,needspace,enumitem,etoolbox,graphicx,physics,microtype,afterpage}
\usepackage{gensymb}

\usepackage[dvipsnames, usenames]{xcolor}

\definecolor{linkcolor}{rgb}{0.0,0.3,0.5}
\usepackage[unicode, colorlinks=true, linkcolor=linkcolor, citecolor=linkcolor, filecolor=linkcolor,urlcolor=linkcolor, pdfusetitle]{hyperref}
\usepackage[all]{hypcap}
\usepackage[T1]{fontenc}
\usepackage[utf8]{inputenc}
\usepackage{tabularx}
\usepackage[normalem]{ulem}

\renewcommand{\arraystretch}{1.4}
\newcommand{\ssim}{\mathchar"5218\relax\,}

\def\apj{\rm{Astrophys.~J.}}
\def\apjl{\rm{Astrophys.~J.~ Lett.}}

\def\mnras{\rm{Mon.~Not.~R.~Astron.~Soc.}}
\def\prd{\rm{Phys.~Rev.~D}}
\def\prl{\rm{Phys.~Rev.~Lett.}}
\def\prx{\rm{Phys.~Rev.~X}}

\def\nat{\rm{Nature}}

\def\cqg{\rm{Class.~Quantum~Gravity}}
\def\lrr{\rm{Living~Rev.~Relativ.}}

\newcommand{\bham}{\affiliation{School of Physics and Astronomy \& Institute for Gravitational Wave Astronomy, \\University of Birmingham, Birmingham, B15 2TT, UK}}
\newcommand{\dallas}{\affiliation{Department of Physics, The University of Texas at Dallas, Richardson, Texas 75080, USA}}

\newcommand\orcid[1]{\href{https://orcid.org/#1}{$\!$\includegraphics[scale=0.006]{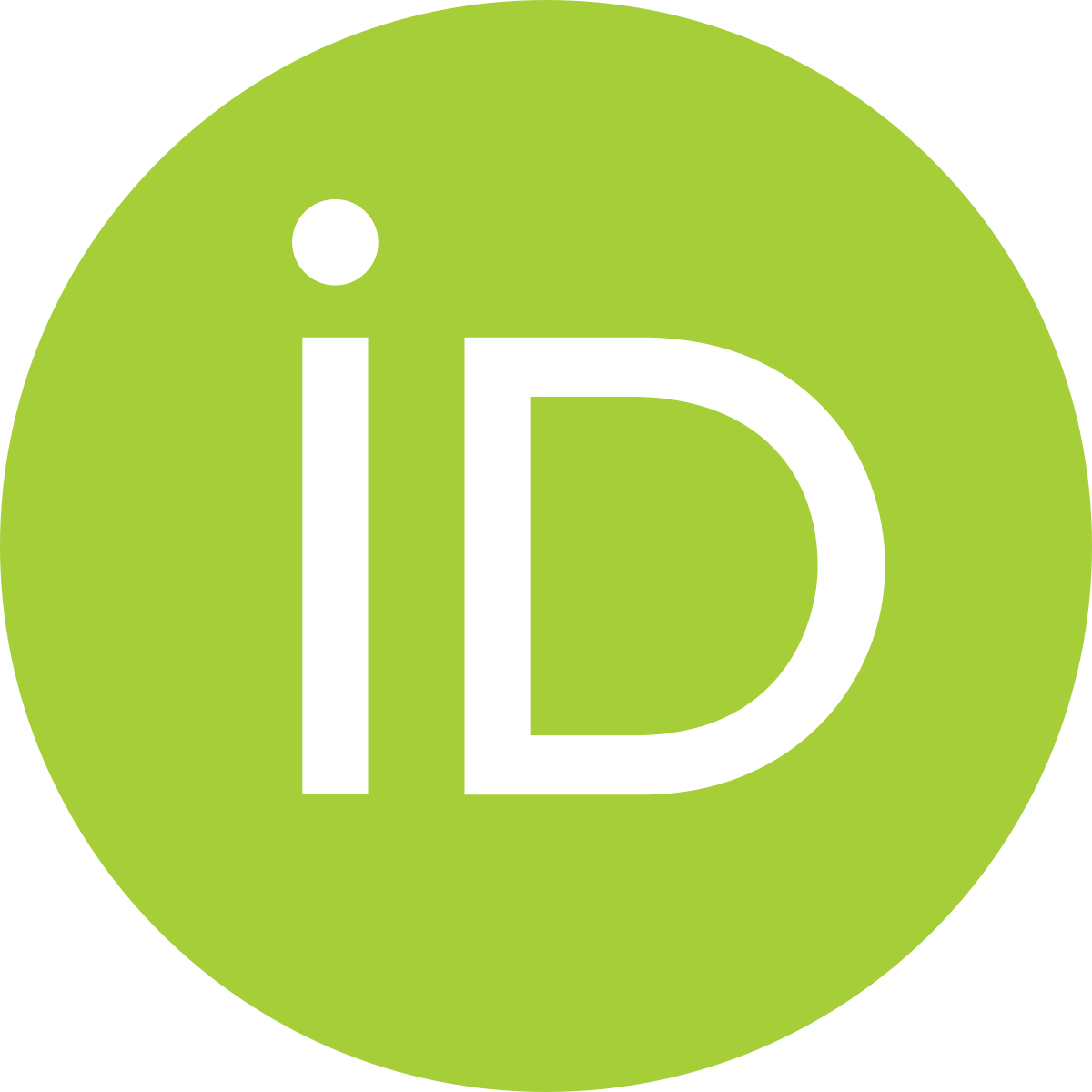} $\!\!$}}

\begin{document}

\title{A taxonomy of black-hole binary spin precession and nutation
}

\author{Daria Gangardt \orcid{0000-0001-7747-689X}}
\thanks{Joint lead author}
\email{ddg672@star.sr.bham.ac.uk}
\bham

\author{Nathan Steinle \orcid{0000-0003-0658-402X}}
\thanks{Joint lead author}
\email{nas161430@utdallas.edu}
\dallas

\author{Michael Kesden \orcid{0000-0002-5987-1471}}
\dallas

\author{Davide Gerosa \orcid{0000-0002-0933-3579}}
\bham

\author{Evangelos Stoikos \orcid{0000-0002-1043-3673}}
\dallas


\date{\today}

\begin{abstract}

Binary black holes with misaligned spins will generically induce both precession and nutation of the orbital angular momentum $\bf{L}$ about the total angular momentum $\bf{J}$.
These phenomena modulate the phase and amplitude of the gravitational waves emitted as the binary inspirals to merger. We introduce a ``taxonomy'' of binary black-hole spin precession that encompasses all the known phenomenology,
then present five new phenomenological parameters that describe generic precession and constitute potential building blocks for future gravitational waveform models.  These are the precession amplitude $\langle\theta_L\rangle$, the precession frequency $\langle \Omega_L\rangle$, the nutation amplitude $\Delta\theta_L$, the nutation frequency $\omega$, and the precession-frequency variation $\Delta\Omega_L$. We investigate the evolution of these five parameters during the inspiral %
and explore their statistical properties for sources with isotropic spins. 
In particular, we find that nutation of $\bf{L}$  is most prominent for binaries with high spins ($\chi \gtrsim 0.5$) and moderate mass ratios ($q \sim 0.6$).

\end{abstract}

\maketitle

\section{Introduction}
\label{sec:Intro}

The LIGO and Virgo gravitational-wave (GW) detectors observed about 50 binary black hole (BBH) mergers %
to date~\cite{2019PhRvX...9c1040A,2020arXiv201014527A,2019ApJ...872..195N,2020PhRvD.101h3030V,2020ApJ...891..123N,2019arXiv191009528Z}. Hundreds (millions) more are expected in the next few years (decades) as current interferometers improve in sensitivity and new detectors begin observations~\cite{2020LRR....23....3A,2010CQGra..27s4002P,2019BAAS...51g..35R,2017arXiv170200786A}. %

Spin precession is a key feature of BBH dynamics \cite{1994PhRvD..49.6274A}.
Although its detectability remains an area of active research \cite{2005ApJ...632.1035O,2020PhRvD.102d1302F,2021PhRvD.103f4067G}, spin precession is expected to be present across the entire black-hole (BH) mass spectrum.
Stellar-mass BBHs ($m \lesssim 100$~M$_{\odot}$) generally originate either from dynamical interactions in dense clusters \cite{2013LRR....16....4B} or from the evolution of isolated binary stars \cite{2014LRR....17....3P}. Measurements of spin precession may help to discriminate relative contributions of each channel to the observed BBH population~\cite{2010CQGra..27k4007M,2013PhRvD..87j4028G,2016PhRvD..93h4029R,2016ApJ...830L..18B,2016PhRvD..94f4020N,2017MNRAS.471.2801S,2017MNRAS.465.4375N,2017CQGra..34cLT01V,2017PhRvD..96b3012T,2017Natur.548..426F,2018PhRvD..97d3014W,2018PhRvD..98h4036G,2018ApJ...854L...9F,2019ApJ...886...25B,2019ApJ...882L..24A,2021ApJ...910..152Z,2020ApJ...895..128M,2020arXiv201109570C}. In particular, dynamically formed BBHs are expected to have isotropically oriented spins and thus present generic precession features, while the spins of BBHs formed in the isolated channel might be more closely aligned with their orbital angular momentum. For supermassive BBHs targeted by the LISA mission, spin precession might help discriminating between the gas-poor and gas-rich nature of their galactic hosts~\cite{2007ApJ...661L.147B,2008ApJ...684..822B,2014ApJ...794..104S,2015MNRAS.451.3941G,2021MNRAS.501.2531S}.

Individual BBH merger events observed by LIGO and Virgo do not currently show unambiguous evidence for spin precession, but population constraints are more promising.  Phenomenological models  find, at 90\% credibility, that half of a BBH population's spins have modest components in the orbital plane \cite{2020arXiv201014533T}. This
is due to the combination of many systems with weak in-plane spin components rather than %
only a few systems with strong precession dynamics.
According to these models, perfect spin alignment for the entire population is excluded at $>99\%$ credibility,
with a preference for the cosines of the spin-orbit misalignment angles to be positive.

The theoretical understanding of BBH spin precession has received much attention in recent decades. As BBHs inspiral towards merger, GW emission is described to leading post-Newtonian (PN) order by the quadrupole formula.  %
Spin-orbit and spin-spin terms couple the binary orbital angular momentum $\mathbf{L}$ and BBH spins $\mathbf{S}_1$ and $\mathbf{S}_2$, modulating the polarization, phase, and amplitude of the emitted GWs \cite{1980RvMP...52..299T,1994PhRvD..49.2658C,1994PhRvD..49.6274A,1995PhRvD..52..821K}. These effects are well understood and form the basis of current data analysis of compact-binary-coalescence GW signals~\cite{2019PhRvD..99f4045V,2020PhRvD.102d4055O,2020arXiv200406503P}. The effects of BBH spins are typically interpreted in terms of parameters such as the projected effective spin $\chi_{\rm eff}$~\cite{2008PhRvD..78d4021R,2018PhRvD..98h3007N} and the precession effective spin $\chi_{\rm p}$ \cite{2015PhRvD..91b4043S,2021PhRvD.103f4067G}.

Furthermore, a variety of configurations where BBH spin dynamics result in peculiar phenomenologies are now known, including transitional precession \cite{1994PhRvD..49.6274A,2017PhRvD..96b4007Z}, spin-orbit resonances \cite{2004PhRvD..70l4020S,2015PhRvD..92f4016G,2018PhRvD..98h3014A}, dynamical instabilities \cite{2015PhRvL.115n1102G,2016PhRvD..93l4074L,2020PhRvD.101l4037M,2021PhRvD.103f4003V}, emergence of new constants of motion \cite{2017CQGra..34f4004G,2021PhRvD.103f4066T}, and large nutation cycles \cite{2016PhRvD..93d4031L,2019CQGra..36j5003G}. %
This paper attempts to incorporate this richness into a single, comprehensive framework and presents five new parameters that encode the most generic features of BBH spin precession.

\begin{figure*}
  \centering
  \includegraphics[width=0.9\textwidth]{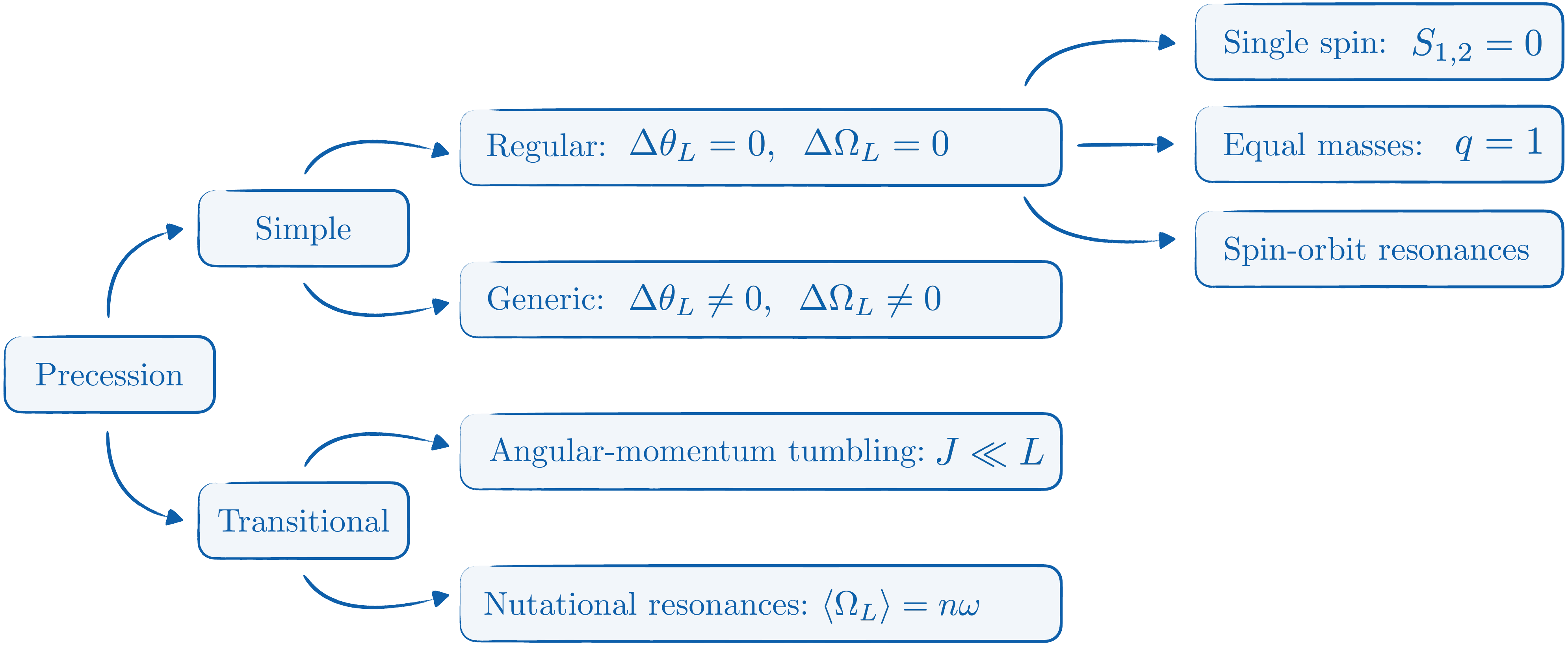}
  \caption{Proposed taxonomy of BBH spin precession. Simple (transitional) precession occurs when the direction of the total angular momentum $\mathbf{J}$ is constant (varying). 
Simple precession is ``regular'' when the orbital angular momentum $\mathbf{L}$ precesses on a cone with fixed opening angle (i.e. $\Delta\theta_L = 0$) and frequency (i.e. $\Delta\Omega_L = 0$) and is ``generic'' when nutation causes the opening angle and frequency to vary (i.e. $\Delta\theta_L \neq 0$, $\Delta\Omega_L \neq 0$). Examples of regular precession include BBHs with a single spin, equal masses, and the spin-orbit resonances of Ref.~\cite{2004PhRvD..70l4020S}. Transitional precession occurs for small values of the total angular momentum ($J \ll L$)~\cite{1994PhRvD..49.6274A} or at nutational resonances ($\langle \Omega_L \rangle = n\omega$) \cite{2017PhRvD..96b4007Z}. 
 }\label{F:Taxonomy}
\end{figure*}

Before delving into the details of this study, we introduce a new ``taxonomy'' of spin precession which encompasses all of the known phenomenology. Our classification is summarized in Fig.~\ref{F:Taxonomy}.
\begin{enumerate}[label={(\arabic*)}]
\item Following \citeauthor{1994PhRvD..49.6274A}~\cite{1994PhRvD..49.6274A},
we refer to precession as ``simple'' when the direction of the total angular momentum $\mathbf{J} = \mathbf{L} + {\mathbf S}_1 + {\mathbf S}_2$ is approximately constant.  In this case, the direction of $\mathbf{L}$ as it precesses about $\mathbf{J}$ can be specified by the polar angle $\theta_L$ and azimuthal angle $\Phi_L$.
\begin{enumerate}%

\item If the total spin magnitude $S = |{\mathbf S}_1 + {\mathbf S}_2|$ is conserved on the precession timescale,
then $\theta_L$ and the precession frequency $\Omega_L \equiv d\Phi_L/dt$ are also constant on this timescale. %
We refer to this uniform precession of $\mathbf{L}$ on a cone about $\mathbf{J}$ as ``regular'' following \citeauthor{1969mech.book.....L}~\cite{1969mech.book.....L}.
Cases of regular precession include:
 
\begin{enumerate}[label={(\roman*)}]

\item a single nonzero spin \cite{1994PhRvD..49.6274A},

\item the equal-mass ($q = 1$) limit \cite{1994PhRvD..49.6274A,2017CQGra..34f4004G},

\item the spin-orbit resonances \cite{2004PhRvD..70l4020S}.

\end{enumerate}

\item In the ``generic'' case when $S$ is \emph{not} constant on the precession timescale~\cite{2015PhRvL.114h1103K,2015PhRvD..92f4016G}, neither are $\theta_L$ nor $\Omega_L$, implying that $\mathbf{L}$ nutates as it precesses about $\mathbf{J}$.

\end{enumerate}

\item %
``Transitional'' precession occurs when the direction of  $\mathbf{J}$ is not constant \cite{1994PhRvD..49.6274A}. There are at least two different but related scenarios when this can occur:

\begin{enumerate}%

\item %
If the ratio of the magnitudes of the total and orbital angular momenta is less than the ratio of the precession and radiation-reaction timescales $J/L \lesssim t_{\rm pre}/t_{\rm RR} \propto (r/M)^{-3/2}$, the direction of $\mathbf{J}$ tumbles \cite{1994PhRvD..49.6274A}.

\item %
At a nutational resonance \cite{2017PhRvD..96b4007Z} where the mean precession frequency is an integer multiple of the nutation frequency (i.e. $\langle \Omega_L \rangle = n\omega$), coherent GW emission tilts the direction of $\mathbf{J}$.

\end{enumerate}
\end{enumerate}

The vast majority of the binaries at a given separation $r$ will undergo generic simple precession (1b), as the other three cases are restricted to finely tuned (2a) or lower-dimensional (1a, 2b) portions of BBH parameter space.  However, as the precession and nutation frequencies evolve during the inspiral, an order-unity fraction of binaries will pass through one or more nutational resonances for comparable mass ratios ($q\lesssim 1$) \cite{2017PhRvD..96b4007Z}.

In this paper, we step back from current GW analyses and waveform models
and attempt to identify those parameters that most naturally characterize the essential %
features of the more common simple precession.
Regular precession (1a) of $\mathbf{L}$ on a cone about $\mathbf{J}$ can be described by the \emph{precession amplitude} $\theta_L$ and the \emph{precession frequency} $\Omega_L$ which are constant on the precession timescale.
However, in the generic case (1b), nonzero nutation implies that the precession amplitude and frequency oscillate about their precession-averaged values $\langle\Omega_L\rangle$ and $\langle\theta_L\rangle$ with \emph{nutation amplitude} $\Delta\theta_L$ and \emph{precession-frequency variation} $\Delta\Omega_L$ at common \emph{nutation frequency} $\omega$.

We stress that nutation is a generic feature of BBH spin dynamics and as such deserves further attention. Our five parameters provide a new framework to characterize configurations in which precession and nutation both significantly impact the dynamics and allow us to isolate and analyze their respective contributions. %
We expect each of these five parameters to imprint a distinct observational signature because of the dominant effect of the direction of $\mathbf{L}$ on the quadrupole waveform \cite{1994PhRvD..49.6274A}, but we leave the characterization of these signatures and the signal-to-noise ratios needed to observe them to future work. 

Our paper is structured as follows. Section~\ref{fivepar} defines and details the five precession parameters we propose. Section~\ref{sec:results} explores their behavior using numerical PN evolutions. Section~\ref{sec:conclusions} summarizes our findings and future prospects. Some details are postponed to Appendices~\ref{sec:Wide_nutations} and \ref{sec: precession frequency diverges}.
We use geometric units where $G=c=1$.

\section{Five precession parameters}
\label{fivepar}

\subsection{Multi-timescale Analysis}
\label{subsec:Multi}

We briefly review the multi-timescale spin-precession framework \cite{2015PhRvL.114h1103K,2015PhRvD..92f4016G} that allows us to calculate our five precession parameters at 2PN order.
BBHs have component masses $m_{1,2}$, mass ratio $q \equiv m_2/m_1\leq 1$, total mass $M \equiv m_1+m_2$, and symmetric mass ratio $\eta \equiv m_1 m_2 / M^2$.  The BBH spins $\mathbf{S}_{1,2}$ determine the dimensionless Kerr parameters $\chi_i \equiv S_i/m_i^2$ and the misalignment angles $\theta_i \equiv \arccos ( \hat{\mathbf{S}}_i \cdot \hat{\mathbf{L}})$, where $\mathbf{L}$ is the Newtonian orbital angular momentum.

BBH evolution occurs on three different timescales:
\begin{enumerate}[label={(\arabic*)}]

\item The binary separation vector $\bf{r}$ changes direction on the orbital timescale $t_{\rm orb}/M \propto (r/M)^{3/2}$.

\item The spins ${\mathbf S}_{1,2}$ and orbital angular momentum $\bf{L}$ precess about the total angular momentum $\bf{J}$ on the precessional timescale $t_{\rm pre}/M \propto (r/M)^{5/2}$.

\item GW emission decreases the binary separation $r$ on the radiation-reaction timescale $t_{\rm RR}/M \propto (r/M)^4$.

\end{enumerate}
The PN approximation ($r\gg M$) implies the timescale hierarchy $t_{\rm orb} \ll t_{\rm pre} \ll t_{\rm RR}$.

The evolution of the three vectors ${\bf S}_1, {\bf S}_2$ and $\bf{L}$ on the precession timescale is a nine-parameter problem, but the seven constraints provided by the constancy of the magnitudes $S_1$, $S_2$, and $L$, the total angular momentum ${\bf J}$, and the projected effective spin \cite{2001PhRvD..64l4013D,2008PhRvD..78d4021R}
\begin{equation} \label{chieffdef}
\chi_{\rm eff} \equiv \left[(1 + q)\mathbf{S}_1 + \left(1 + \frac{1}{q}\right)\mathbf{S}_2 \right] \cdot \frac{\hat{\mathbf{L}}}{M^2}
\end{equation}
leave only two degrees of freedom.  We choose the total spin magnitude $S$ and the azimuthal angle $\Phi_L$ as generalized coordinates to describe these degrees of freedom; $S$ is an intrinsic parameter, while $\Phi_L$ is extrinsic.

The spin magnitude $S$ oscillates between turning points $S_\pm$ with a period
\begin{equation} \label{E:Tau}
    \tau = 2\int_{S_{-}}^{S_{+}}\frac{dS}{|dS/dt|}\,,
\end{equation}
where the turning points $S_\pm$ are functions of $S$ and the constants of motion, 
\begin{align} \label{E:dSdt}
\frac{dS}{dt} &= -\frac{3}{2} \left( \frac{1 - q}{1 + q} \right) \frac{S_1S_2}{MS} \left[ 1 - \chi_{\rm eff} \left( \frac{M}{r} \right)^{1/2} \right] \left( \frac{r}{M} \right)^{-5/2}
\notag \\
& \times \sin\theta_1 \sin\theta_2 \sin\Delta\Phi_{12}\,,
\end{align}
and $\Delta\Phi_{12}$ is the angle between the components of ${\bf S}_1$ and ${\bf S}_2$ in the orbital plane.  As $S$ is the only intrinsic parameter, we can define the precession average of any quantity $X(S)$ by:
\begin{equation} \label{E:PrecAve}
\langle X\rangle \equiv \frac{2}{\tau} \int_{S_-}^{S_+} \frac{X(S)}{|dS/dt|} dS\,.
\end{equation}
Let us also define $X_\pm \equiv X(S_\pm)$ for any quantity $X$.

The angular momentum $\bf{L}$ precesses about $\bf{J}$ with precession frequency
\begin{align} \label{E:Omegaz}
\begin{aligned}
\Omega_L(S) &= \frac{d\Phi_L}{dt}
= \frac{J}{2r^3} %
\Bigg( 1 + \frac{3(1+q)}{2q} \left[ 1 - \chi_{\rm eff} \left( \frac{M}{r} \right)^{1/2} \right]
 \\
& \times \bigg\{ 1 + q - [J^2 - (L - S)^2]^{-1}[(L + S)^2 - J^2]^{-1} \\
& \times \big[4(1-q)L^2(S_1^2 - S_2^2) - (1+q)(J^2 - L^2 -S^2) \\
& \times (J^2 - L^2 -S^2 - 4\eta M^2L\chi_{\rm eff}) \big] \bigg\} \Bigg)\,,
\end{aligned}
\end{align}
During a period $\tau$,  $\bf{L}$ precesses by an angle $\alpha = \langle \Omega_L \rangle \tau$ about $\bf{J}$.

As the polar angle between $\mathbf{L}$ and $\mathbf{J}$ is given by
\begin{align}\label{E:ThetaL}
\theta_L(S) = \arccos \left( \frac{J^2 + L^2 - S^2}{2JL} \right)\,,
\end{align}
the oscillation of $S$ with period $\tau$ leads to nutation of $\mathbf{L}$ with frequency $\omega \equiv 2\pi/\tau$.

\subsection{Precession parameters}
\label{subsec:Param}

This formalism highlights five promising parameters to describe simple precession:
\begin{enumerate}[label={(\arabic*)}]
    \item The precession amplitude given by the average $\langle \theta_L \rangle$ or median $\overline{\theta}_L \equiv (\theta_{L+} + \theta_{L-})/2$\,.

    \item The precession frequency given by the average $\langle\Omega_L\rangle $ or median $\overline{\Omega}_L \equiv (\Omega_{L+} + \Omega_{L-})/2$\,.

    \item The nutation amplitude $\Delta\theta_L \equiv (\theta_{L+} - \theta_{L-})/2$\,. %

    \item The nutation frequency $\omega \equiv 2\pi/\tau$\,.

    \item The precession-frequency variation $\Delta\Omega_L \equiv (\Omega_{L+} - \Omega_{L-})/2$\,.
\end{enumerate}
The nutation amplitude $\Delta\theta_L$ and precession-frequency variation $\Delta\Omega_L$ vanish for regular precession (1a); $\theta_L$ and $\Omega_L$ oscillate with the same nutation frequency $\omega$ because $S$ is the only intrinsic parameter varying on the precession timescale.

\subsection{Leading PN behavior}
\label{leadingPN}

We can develop intuition about our five precession parameters by calculating their values at leading PN order ($r/M \to \infty$).  In this limit, 1.5PN spin-orbit coupling dominates over 2PN spin-spin coupling \cite{1995PhRvD..52..821K}.  The individual spins $\mathbf{S}_{1,2}$ precess regularly on cones about the orbital angular momentum $\mathbf{L}$ with opening angles $\theta_{1\infty}$ and $\theta_{2\infty}$ and frequencies
\begin{eqnarray}
\Omega_{1\infty} &=& \frac{(4 + 3q)\eta}{2M} \left( \frac{r}{M} \right)^{-5/2}\,,  \label{E:Omega1Inf} \\
\Omega_{2\infty} &=& \frac{(4 + 3/q)\eta}{2M} \left( \frac{r}{M} \right)^{-5/2}\,. \label{E:Omega2Inf}
\end{eqnarray}
Defining $\mathbf{X}_\perp$ as the component of vector $\mathbf{X}$ perpendicular to the total angular momentum $\mathbf{J}$, the precession amplitude $\langle \theta_L \rangle$ in the limit $r/M \to \infty$ is
\begin{eqnarray}
\langle \theta_L \rangle_\infty &=& \frac{\langle S_\perp \rangle}{L} \notag\\
&=& \left\{ \left[ \left( \frac{\chi_1 \sin\theta_{1\infty}}{q} \right)^2\!\! + \!(q\chi_2 \sin\theta_{2\infty})^2 \right] \frac{M}{r} \right\}^{1/2} \!\!\!\!\!\!\!.~~~\label{E:ThetaLInf}
\end{eqnarray}
The %
precession frequency $\langle \Omega_L \rangle$ is bimodal in the limit $r/M \to \infty$ and given by
\begin{eqnarray}
\langle \Omega_L \rangle_\infty =
\begin{cases}
 \Omega_{1\infty} \qquad {\rm if} \quad S_{1\perp} > S_{2\perp}\,, \\
\Omega_{2\infty} \qquad {\rm if} \quad S_{1\perp} < S_{2\perp}\,.
\end{cases}
 \label{E:OmegaInf}
\end{eqnarray}
This result, expressed in a different notation, was first presented in Eqs.~(46) and (47) of Ref.~\cite{2017PhRvD..96b4007Z}.  For a population of BBHs with given values of $q,\chi_1,\chi_2$ and isotropic spin directions,
the fraction of sources with $S_{1\perp} < S_{2\perp}$ is given by \cite{2017PhRvD..96b4007Z}
\begin{eqnarray}
f_< =
\begin{cases}
\displaystyle \frac{|\chi_1^2 - q^4\chi_2^2|}{4q^2\chi_1\chi_2} (\sinh H_C - H_C) & {\rm if}\quad S_1 > S_2,\\
\displaystyle \frac{|\chi_1^2 - q^4\chi_2^2|}{4q^2\chi_1\chi_2} (\sinh H_S + H_S) & {\rm if}\quad  S_1 < S_2,
\end{cases}~~~
 \label{E:fsmaller}
\end{eqnarray}
where
\begin{eqnarray}
H_C &=& 2\cosh^{-1}\left( \frac{\chi_1}{|\chi_1^2 - q^4\chi_2^2|^{1/2}} \right)\,, \\
H_S &=& 2\sinh^{-1}\left( \frac{\chi_1}{|\chi_1^2 - q^4\chi_2^2|^{1/2}} \right)\,.
\end{eqnarray}

The nutation amplitude $\Delta\theta_L$ in the limit $r/M \to \infty$ is similarly bimodal and given by
\begin{eqnarray}
 \label{E:DeltaThetaLInf}
\Delta\theta_{L\infty} &=& \frac{1}{2L} (S_{1\perp} + S_{2\perp} - |S_{1\perp} - S_{2\perp}|)
\notag \\
&=& \begin{cases}
\displaystyle q\chi_2\sin\theta_{2\infty} \left( \frac{M}{r} \right)^{1/2} & {\rm if} \quad  S_{1\perp} > S_{2\perp}\,, \\
\displaystyle \frac{\chi_1}{q} \sin\theta_{1\infty} \left( \frac{M}{r} \right)^{1/2}  & {\rm if} \quad  S_{1\perp} < S_{2\perp}\,.
 \end{cases}~~~
\end{eqnarray}

The nutation frequency $\omega$ in the limit $r/M \to \infty$ is
\begin{equation} \label{E:omegaInf}
\omega_\infty = \Omega_{2\infty} - \Omega_{1\infty} = \frac{3}{2M} \left( \frac{1 - q}{1 + q} \right) \left( \frac{r}{M} \right)^{-5/2}\,,
\end{equation}
which is independent of the BBH spins and vanishes in the equal-mass limit $q \to 1$, consistent with the constancy of $S$ in this limit even at 2PN order \cite{2017CQGra..34f4004G}.  This implies that $\theta_L$ and $\Omega_L$ are also constant and that Eqs.~(\ref{E:ThetaLInf}), (\ref{E:DeltaThetaLInf}), and (\ref{E:OmegaPMInf}) are invalid in the precisely equal-mass limit. 

The precession frequency $\Omega_{L \pm}$ at $S = S_\pm$ in the limit $r/M \to \infty$ is
\begin{eqnarray}
\Omega_{L \pm\infty} %
&=& \frac{\chi_1 \sin\theta_{1\infty}\Omega_{1\infty} \pm q^2\chi_2 \sin\theta_{2\infty}\Omega_{2\infty}}{\chi_1 \sin\theta_{1\infty} \pm q^2\chi_2 \sin\theta_{2\infty}}\,, \label{E:OmegaPMInf}
\end{eqnarray}
implying that the precession-frequency variation $\Delta\Omega_L$ in the limit $r/M \to \infty$ is
\begin{eqnarray}
\Delta\Omega_{L\infty} &=& \frac{1}{2}(\Omega_{L+\infty} - \Omega_{L-\infty}) \notag \\
&=& \frac{q^2 \chi_1 \sin\theta_{1\infty} \chi_2 \sin\theta_{2\infty}}{\chi_1^2 \sin^2\theta_{1\infty} - q^4 \chi_2^2 \sin^2\theta_{2\infty}} \omega_\infty\,.
\label{E:DeltaOmegaInfty}
\end{eqnarray}

\section{Numerical evolutions}
\label{sec:results}

We now explore the evolution and distribution of our five precession parameters. Numerical integrations are performed with the \textsc{precession} code~\cite{2016PhRvD..93l4066G}, which implements 2PN spin-precession equations~\cite{2008PhRvD..78d4021R} and 1.5 PN precession-averaged radiation reaction~\cite{2015PhRvL.114h1103K,2015PhRvD..92f4016G}. Sources are evolved from their asymptotic conditions at $r/M \to \infty$
down to $r=10M$, taken as the threshold for the breakdown of the PN approximation.

\subsection{Individual sources}
\label{indsources}

\begin{figure*}[!t]
  \centering
  \includegraphics[width=\textwidth]{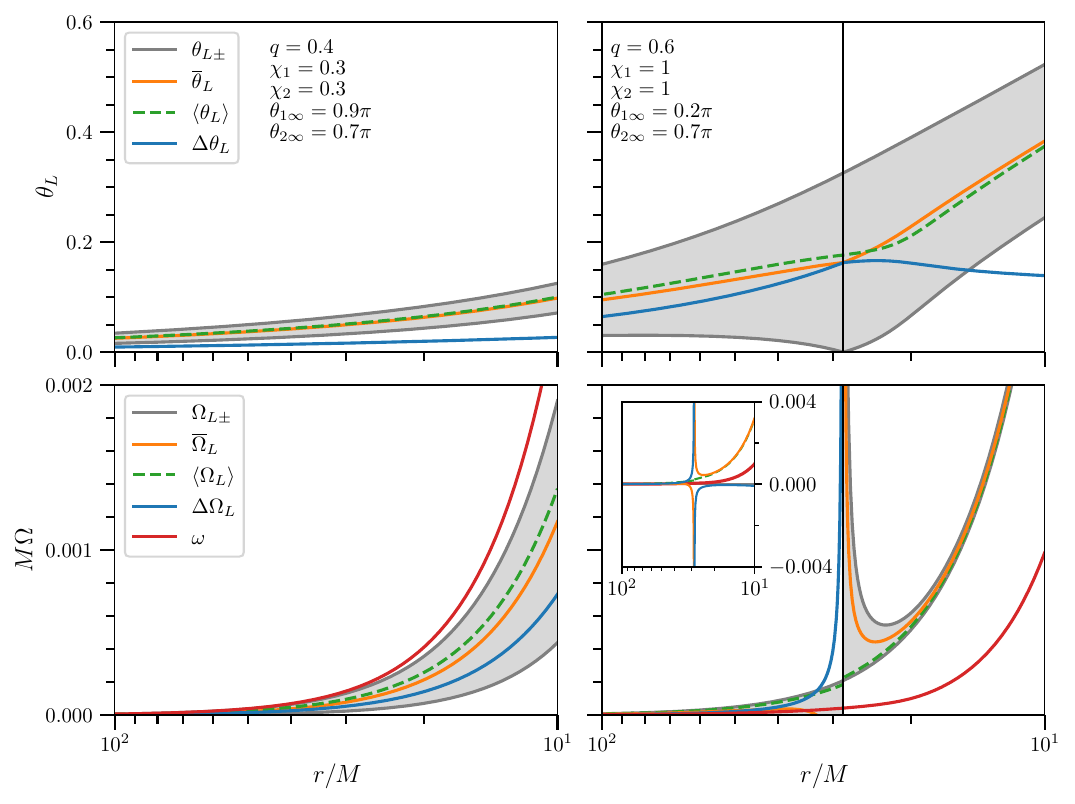}
   \caption{Evolution of our five precession parameters for two representative inspirals. These are
 the precession amplitude $\langle\theta_L\rangle$ (dashed green, top),
 the nutation amplitude $\Delta\theta_L$ (solid blue; top),
 the precession frequency $\langle\Omega_L\rangle$ (dashed green, bottom),
 the precession-frequency variation $\Delta\Omega_L$ (solid blue, bottom), and the nutation frequency $\omega$ (solid red, bottom). We also show the medians $\overline{\theta}_L$ and $\overline{\Omega}_L$ (solid orange), as well as the allowed ranges ${\theta}_{L\pm}$ and ${\Omega}_{L\pm}$ (gray curves and shaded areas). The two binaries shown in this figure are characterized by the values of $q,\chi_1,\chi_2,\theta_{1\infty},$ and $\theta_{2\infty}$ listed in the top panels. The right (left) panels depict a case where $\sin\theta_{L}$ does (not) reach $0$ at some point during the inspiral. This condition is marked by a vertical black line ($r \approx 27 M$ for the binary on the right). %
   }\label{F:Evolutions}
\end{figure*}

Figure~\ref{F:Evolutions} %
displays two representative cases for the evolution of our five parameters as functions of the separation $r$. The key difference between these two %
systems is whether $\bf{J}$ and $\bf{L}$ align at some point during the inspiral. Because the function $\theta_L(S)$ given by Eq.~(\ref{E:ThetaL}) is monotonic, the condition $\sin\theta_{L}(S)=0$ can only be satisfied if either $S_- = |J-L|$ or $S_+ = J+L$,  which correspond to $\theta_{L-}=0$ and $\theta_{L+}=\pi$, respectively~\cite{2017PhRvD..96b4007Z}.  Appendix~\ref{sec:Wide_nutations} shows our proof that these two conditions cannot be satisfied simultaneously, i.e. maximal nutations $\Delta\theta_L = \pi$ are forbidden. This is unlike nutations of ${\bf S}_1$ and ${\bf S}_1$ which can have maximal amplitude $\pi$ during a single period $\tau$~\cite{2019CQGra..36j5003G}. %

The left panels of Fig.~\ref{F:Evolutions} show a binary for which $\theta_{L}$ never reaches $0$ or $\pi$.
The average precession and nutation amplitudes $\langle \theta_L \rangle$ and $\Delta\theta_L$ are approximately proportional to $(r/M)^{-1/2}$ as suggested by the leading-order behavior given by Eqs.~(\ref{E:ThetaLInf}) and (\ref{E:DeltaThetaLInf}), while the three frequencies $\langle \Omega_L \rangle$, $\omega$, and $\Delta\Omega_L$ are nearly proportional to $(r/M)^{-5/2}$ consistent with the leading-order behavior given by Eqs.~(\ref{E:OmegaInf}) and (\ref{E:omegaInf}), and (\ref{E:DeltaOmegaInfty}).  The two precession averages $\langle \theta_L \rangle$ and $\langle \Omega_L \rangle$ are well approximated by the median values $\overline{\theta}_L$ and $\overline{\Omega}_L$, as one would expect at small nutation amplitude $\Delta\theta_L$ where the oscillations are nearly sinusoidal.

The evolution of our five precession parameters is somewhat more complex if $\mathbf{L}$ and $\mathbf{J}$ reach co-alignment at some point during the inspiral. The right panels of Fig.~\ref{F:Evolutions} show an example of such a binary where a cusp-like minimum $\theta_{L-}=0$
and a corresponding cusp-like local maximum in the nutation amplitude $\Delta\theta_L$ occur at $r \approx 27 M$.
If $\boldsymbol{\Omega}$ is the precession vector, i.e. $d\mathbf{L}/dt = \boldsymbol{\Omega} \times \mathbf{L}$, then the precession frequency of Eq.~(\ref{E:Omegaz}) is
\begin{equation} \label{E:OmegaLgeo}
\Omega_L = \frac{d\mathbf{L}}{dt} \cdot \frac{\hat{\mathbf{J}} \times \hat{\mathbf{L}}_\perp}{L_\perp} = \boldsymbol{\Omega} \cdot (\hat{\mathbf{J}} - \hat{\mathbf{L}}_\perp \cot\theta_L)\,.
\end{equation}
In Appendix~\ref{sec: precession frequency diverges}, we show that $\boldsymbol{\Omega} \cdot \hat{\mathbf{L}}_\perp \neq 0$ for misaligned spins,
implying that
the second term in Eq.~(\ref{E:OmegaLgeo}) diverges and thus $\Omega_{L-}$ approaches $\pm\infty$ as $\theta_{L-}$ approaches zero (or $\theta_{L+}$ approaches $\pi$) during the inspiral.  As $\mathbf{L}$ passes through alignment with $\mathbf{J}$, $\hat{\mathbf{L}}_\perp \to -\hat{\mathbf{L}}_\perp$ and $\Omega_{L-}$ goes to $\mp\infty$ according to Eq.~(\ref{E:OmegaLgeo}).  This can be seen in the bottom right panel of Fig.~\ref{F:Evolutions}, where $\Omega_{L-}$ jumps from $-\infty$ to $+\infty$ as the binary inspirals through $r \approx 27 M$
at which $\theta_{L-} = 0$.
The precession-frequency variation $\Delta\Omega_L$ correspondingly jumps from $+\infty$ to $-\infty$.  Integrating Eq.~(\ref{E:OmegaLgeo}) with respect to time, we find that this discontinuity causes the precession angle per nutation period $\alpha \equiv \int_0^\tau \Omega_L~dt$  to jump by $\Delta\alpha = \pm 2\pi$ and the average precession frequency $\langle\Omega_L\rangle = \alpha/\tau$ to jump by $\Delta\langle \Omega_L \rangle = \pm\omega$ as explored by  \citeauthor{2017PhRvD..96b4007Z}~\cite{2017PhRvD..96b4007Z}.  Careful examination of the dashed green curve in the bottom right panel of Fig.~\ref{F:Evolutions} reveals this discontinuity in $\langle\Omega_L\rangle$ at the vertical black line.  Numerical exploration did not reveal binaries with two or more of such $\mathbf{L}\parallel\mathbf{J}$ crossings.

\subsection{Parameter-space exploration}

\begin{figure*}[!t]
  \centering
  \includegraphics[width=\textwidth]{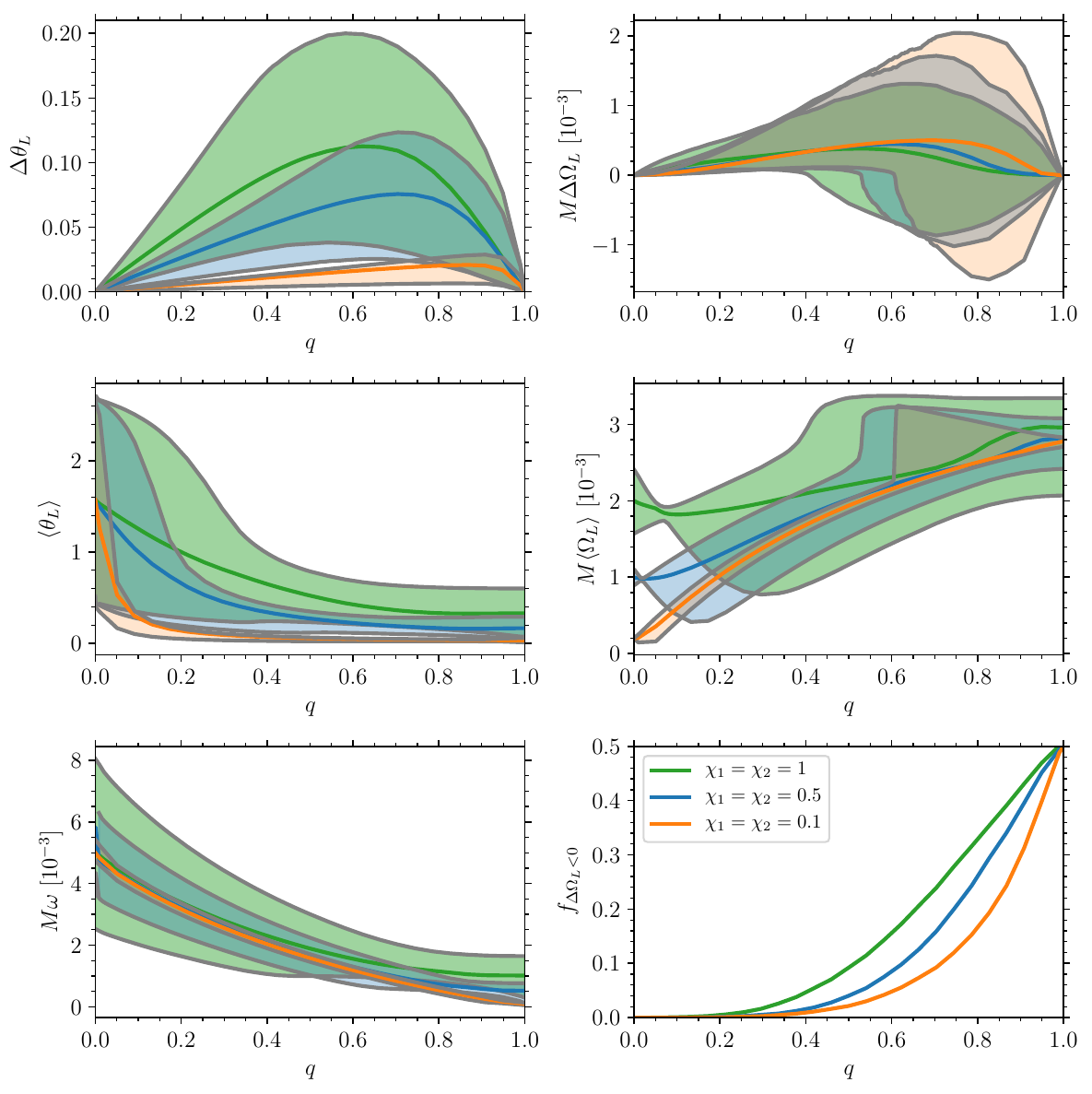}
   \caption{Distributions of the precession parameters $\Delta\theta_L$ (top left), $\langle \theta_L \rangle$ (middle left), $\omega$ (bottom left), $\Delta\Omega_L$ (top right), and $\langle \Omega_L \rangle$ (middle right) as functions of the mass ratio $q$ for isotropic distributions of spin directions at $r = 10M$. The solid orange, blue, and green lines show the median values for spin magnitudes $\chi_1 = \chi_2 = 0.1,~0.5,$ and $1$, while the shaded areas indicate the 90\% interval of each distribution.  The bottom right panel shows the fraction of binaries with $\Delta\Omega_L < 0$ for the same BBHs.
 }\label{F:Contours}
\end{figure*}

The dependence of the five precession parameters at $r=10 M$ on the mass ratio $q$ is shown in Fig.~\ref{F:Contours} for three values of the spin magnitudes $\chi_1 = \chi_2 = 1,~0.5,$ and $0.1$.
The binaries for which the nutation amplitude $\Delta\theta_L$ is largest have high spins
but moderate mass ratio $q \sim 0.6$. This counterintuitive result constitutes one of the key findings of this paper. Two-spin effects are, naively, maximized for comparable-mass sources $q\lesssim 1$ because the secondary's spin $S_2$ vanishes for $q\to 0$. This is \emph{not} the case for nutations. The magnitude $S$ becomes a constant of motion in both the $q\to 0$ and the $q\to 1$ limits, which implies $\Delta\theta_L=\Delta\Omega_L=0$. Nutation effects are set by the variation of $S$ and are more prominent for binaries with moderate mass ratios.

As expected, large values of $\Delta\theta_L$ are more likely for high $\chi_1$ and $\chi_2$, because large spins can induce greater misalignments between the total and orbital angular momenta (i.e. $\mathbf{J} - \mathbf{L} = \mathbf{S}_1 + \mathbf{S}_2$).
Figure~\ref{F:Contours} also shows that the maximum value of $\Delta\theta_L$ occurs at smaller $q$ if $\chi_1 = \chi_2$ increases.
This can be understood in terms of the spin-precession morphologies explored in depth in Refs.~\cite{2015PhRvL.114h1103K,2015PhRvD..92f4016G}. Nutation is larger in the circulating morphology than the two librating morphologies in which the spins merely oscillate about the spin-orbit resonances (case 1a.iii of regular precession in our taxonomy detailed in Sec.~\ref{sec:Intro}).
To maximize the nutation amplitude $\Delta\theta_L$ at higher $\chi_i$, the mass ratio $q$ must decrease to suppress spin-spin coupling and maintain a large fraction of binaries in the circulating morphology.

The precession-frequency variation $\Delta\Omega_L$ also reaches its largest values at moderate mass ratios.  However, unlike for the nutation amplitude $\Delta\theta_L$, smaller spins produce larger variations $\Delta\Omega_L$.  Comparing the leading PN behavior given by Eqs.~(\ref{E:DeltaThetaLInf}) and (\ref{E:DeltaOmegaInfty}), we see that $\Delta\theta_{L\infty}$ is linear in the spin magnitudes, while $\Delta\Omega_{L\infty}$ only depends on their ratio $\chi_2/\chi_1$.  This ratio is unity for all three spin distributions in Fig.~\ref{F:Contours}, but the weaker spin-spin coupling for smaller $\chi_i$ again implies a higher fraction of binaries in the circulating morphology and thus larger variations $\Delta\Omega_L$. The sharp decreases in the lower boundaries of the shaded regions (the $5^{\rm th}$ percentile of each distribution) approximately occur at the values of $q$ at which the fraction of binaries with $\Delta\Omega_L < 0$ reaches 0.05 ($f_{\Delta\Omega_L < 0} = 0.05$ in the bottom right panel of Fig.~\ref{F:Contours}).  In the limit that spin-spin coupling is suppressed, this occurs at $q \simeq 0.62$ where $f_< = 0.05$ according to Eq.~(\ref{E:fsmaller}).

This fraction $f_{\Delta\Omega_L < 0}$ increases with $q$ for $\chi_1=\chi_2$, consistent with the leading PN behavior given by Eq~(\ref{E:DeltaOmegaInfty}).  This equation also shows that in the equal-mass limit $q\to 1$, $\Delta\Omega_{L\infty}$ is equally likely to be positive or negative, consistent with our numerical result that $f_{\Delta\Omega_L < 0} \to 0.5$ in this limit.
The fraction $f_{\Delta\Omega_L < 0}$ is not necessarily maximized at $q = 1$ for $\chi_1 \neq \chi_2$.  For example, we find that $f_{\Delta\Omega_L < 0}$ reaches a maximum of $\ssim0.95$ at $q \simeq 0.65$ 
for $\chi_1 = 0.1$ and $\chi_2 = 1$.

The precession amplitude $\langle\theta_L\rangle$ shown in the middle left panel of Fig.~\ref{F:Contours} decreases monotonically with $q$ and increases monotonically with $\chi_i$.  In the extreme mass-ratio limit $q \to 0$, $\mathbf{L} \to 0$ and $\mathbf{S} \to \mathbf{S}_1$, implying that $\langle\theta_L\rangle \to \pi/2$ for isotropic spin distributions.
As $q$ increases, the two spins can more effectively cancel each other in the vector sum $\mathbf{S} = \mathbf{S}_1 + \mathbf{S}_2$ leading to smaller precession amplitudes $\theta_L$ by Eq.~(\ref{E:ThetaL}).  Larger spin magnitudes lead to larger precession amplitudes both geometrically by Eq.~(\ref{E:ThetaL}) and because enhanced spin-spin coupling increases the fraction of binaries in the precession morphology in which the components of the spins in the orbital plane librate about alignment and thus add constructively.

The average precession frequency $\langle\Omega_L\rangle$ is shown in the middle right panel of Fig.~\ref{F:Contours}.  In the extreme mass-ratio limit $q \to 0$, $\mathbf{J} \to \mathbf{S}_1$ and therefore $\Omega_L \propto \chi_1$ according to Eq.~(\ref{E:Omegaz}).  The larger scatter in the distributions for larger spins in this limit follows from the dependence of $\Omega_L$ on the projected effective spin $\chi_{\rm eff}$ in this equation which spans a larger range $-\chi_1 \leq \chi_{\rm eff} \leq +\chi_1$ for higher spins. %
As $q$ increases, $\langle\Omega_L\rangle$ generally increases as well, particularly for small spins where the leading PN approximation of Eq.~(\ref{E:OmegaInf}) is more accurate.  The sharp increases in the upper boundaries of the shaded regions (the $95^{\rm th}$ percentile of each distribution) are essentially the mirror image of the similar features in the lower boundaries of the $\Delta\Omega_L$ distributions in the top right panel.  This follows from the definitions of these parameters: $\Delta\Omega_L \equiv (\Omega_{L+} - \Omega_{L-})/2$ and $\langle\Omega_L\rangle \approx \overline{\Omega}_L \equiv (\Omega_{L+} + \Omega_{L-})/2$.

The nutation frequency $\omega$ decreases monotonically with $q$, consistent with the factor of $(1-q)/(1+q)$ in Eqs.~(\ref{E:dSdt}) and (\ref{E:omegaInf}).  Its median value is largely independent of the spin magnitude, also consistent with Eqs.~(\ref{E:Tau}) and (\ref{E:omegaInf}).  The widths of the $\omega$ distributions are roughly proportional to $\chi_i$, which follows from the term proportional to $\chi_{\rm eff}$ in Eq.~(\ref{E:dSdt}), similar to the scatter in $\langle\Omega_L\rangle$ in the extreme mass-ratio limit. %

\begin{figure*}[!t]
  \centering
  \includegraphics[width=\textwidth]{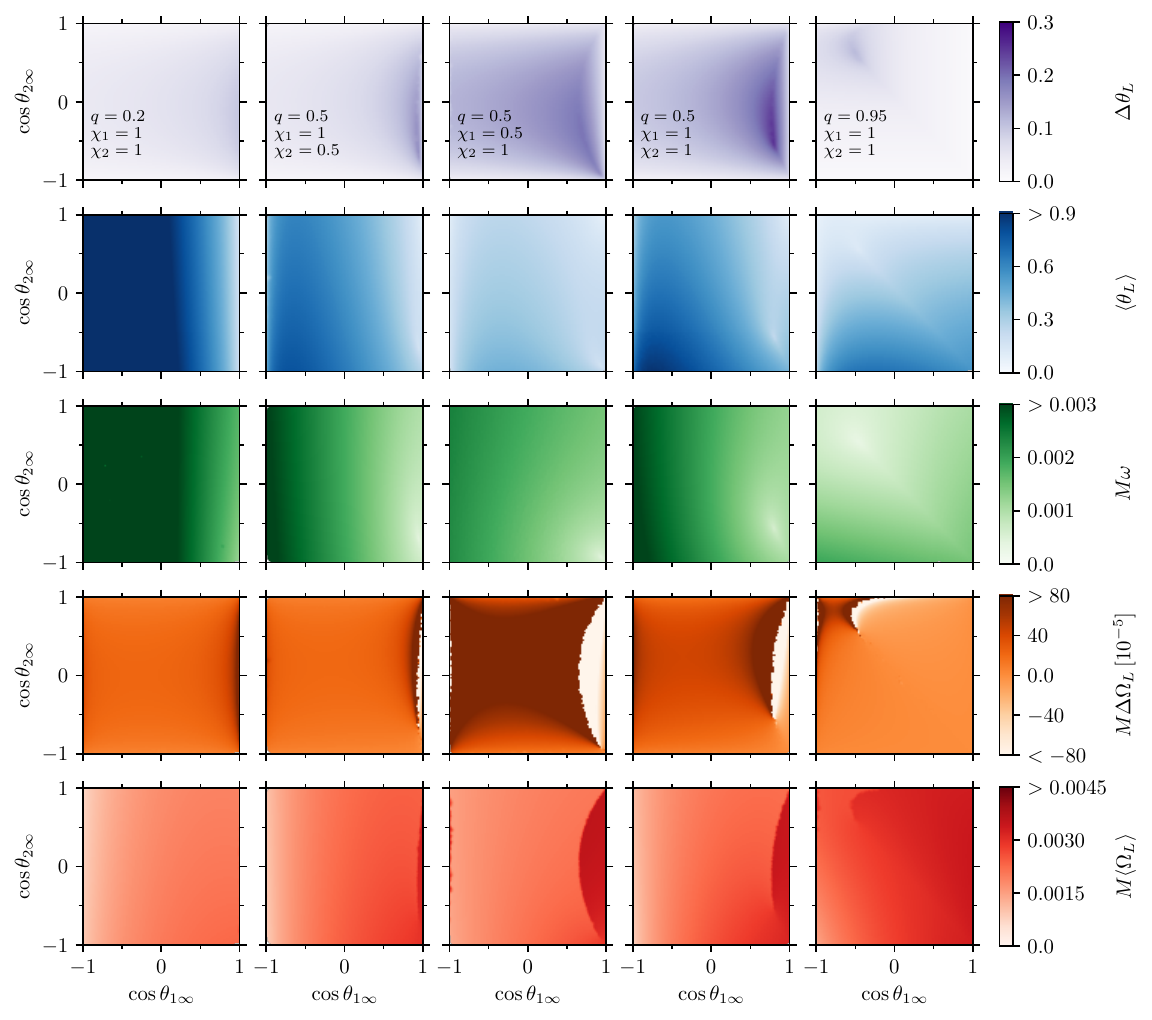}
   \caption{Precession parameters $\Delta\theta_L$, $\langle \theta_L \rangle$, $\omega$, $\Delta\Omega_L$, and $\langle \theta_L \rangle$ (top to bottom) at $r = 10M$ as a function of the asymptotic spin misalignment angles $\theta_{1\infty}$ and $\theta_{2\infty}$. Each column corresponds to a set of values of mass ratio $q$ and spin magnitudes $\chi_i$.  For visualization purposes, the shading saturates above and below the thresholds indicated in the color bars.
   }
 \label{F:5x5}
\end{figure*}

In Fig.~\ref{F:5x5}, we explore how our five precession parameters at $r = 10M$ depend on spin orientation for five choices of mass ratio $q$ and spin magnitudes $\chi_1$ and $\chi_2$.  We parametrize the spin orientations by the cosines of the misalignment angles $\cos\theta_{1\infty}$ and $\cos\theta_{2\infty}$ in the limit $r \to \infty$; these parameters fully determine $J$ and $\chi_{\rm eff}$ at all separations as discussed in Ref.~\cite{2015PhRvD..92f4016G} and can thus be used to calculate the precession parameters as described in Sec.~\ref{fivepar}.  Isotropic spin distributions remain isotropic as they inspiral \cite{2007ApJ...661L.147B} and are thus specified by flat distributions of $\cos\theta_{1\infty}$ and $\cos\theta_{2\infty}$.

The top row of Fig.~\ref{F:5x5} shows the nutation amplitude $\Delta\theta_L$.  The boundaries of the plane ($\cos\theta_{1\infty} = \pm 1, \cos\theta_{2\infty} = \pm 1$) correspond to the spin-orbit resonances \cite{2004PhRvD..70l4020S} that undergo regular precession for which $\Delta\theta_L = 0$  (case 1a.iii of our taxonomy). %
The nutation amplitude increases as one moves inwards from the boundaries and is largest for the three distributions with $q = 0.5$, consistent with Fig.~\ref{F:Contours}.  All three of these distributions possess distinct crests of large $\Delta\theta_L$ that extend from near the bottom right corner of each plot to the top right corner.  A line tracing along this crest corresponds to the set of binaries with $\theta_{L-} = 0$ at $r = 10M$; by minimizing $\theta_{L-}$, these binaries naturally have large values of the nutation amplitude $\Delta\theta_L \equiv (\theta_{L+} - \theta_{L-})/2$ (corresponding to the local maximum of the solid blue curve in the top right panel of Fig.~\ref{F:Evolutions}).  We address the consequences of the condition $\mathbf{J} \parallel \mathbf{L}$ in greater detail in
Sec.~\ref{JparL}.

The second row of Fig.~\ref{F:5x5} shows the precession amplitude $\langle \theta_L \rangle$.  These plots appear anti-correlated with those in the first row, a consequence of the contribution of $\theta_{L-}$ to these parameters: $\Delta\theta_L \equiv (\theta_{L+} - \theta_{L-})/2$ and $\langle \theta_L\rangle \approx \overline{\theta}_L \equiv (\theta_{L+} + \theta_{L-})/2$.  The alternating constructive and destructive addition in the vector sum $\mathbf{S} = \mathbf{S}_1 + \mathbf{S}_2$ that maximizes $\Delta\theta_L$ suppresses the precession-averaged $\langle \theta_L\rangle$.

The precession frequency $\omega$ shown in the third row of Fig.~\ref{F:5x5} has the weakest dependence on spin orientation, consistent with the spin-independent leading-order PN result of Eq.~(\ref{E:omegaInf}).  The higher-order dependence on spin orientation can be largely explained through the term proportional to $\chi_{\rm eff}$ in Eq.~(\ref{E:dSdt}).  Another feature of these plots, also apparent in the second row, is the weak dependence on $\cos\theta_{2\infty}$ for small mass ratio $q$ or $\chi_2 \ll \chi_1$.

\begin{figure*}[!t]
  \centering
  \includegraphics[width=\textwidth]{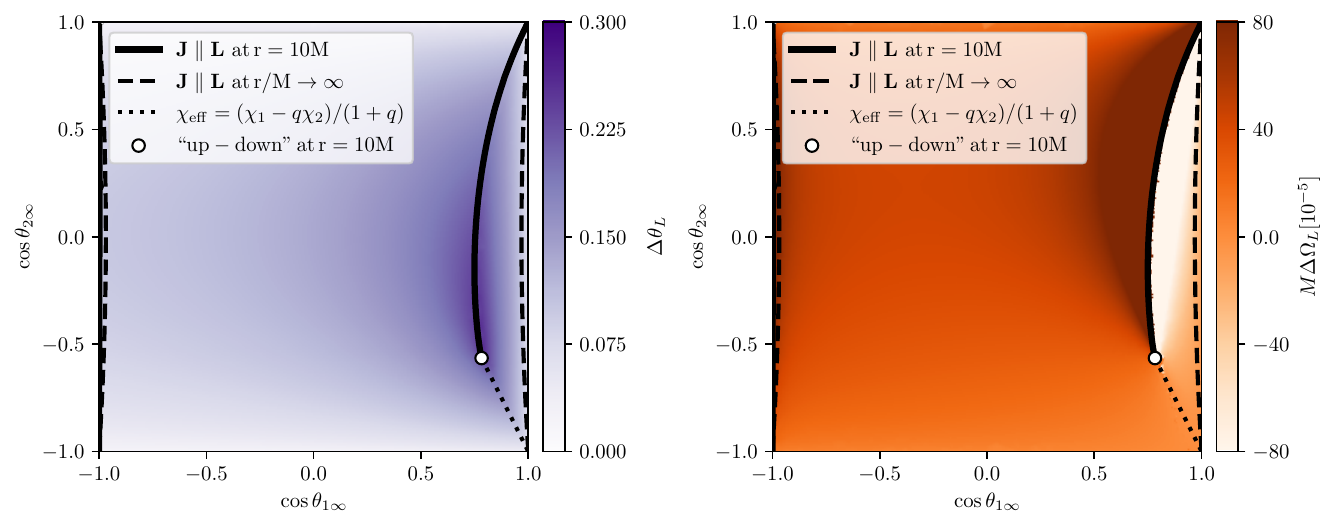}
   \caption{The nutation amplitude $\Delta\theta_L$ and precession-frequency variation $\Delta\Omega_L$ at $r=10M$ as functions of the cosines of the asymptotic misalignment angles $\cos{\theta_{1\infty}}$ and $\cos{\theta_{2\infty}}$ for $q=0.5$ and $\chi_1=\chi_2=1$. The solid (dashed) black lines depict the asymptotic origin of binaries which are found with $\bf{J} \parallel \bf{L}$ at $r=10 M$ ($r/M\to \infty$).
   Binaries that precess through the unstable ``up-down'' configuration (i.e. $\cos\theta_1=-\cos\theta_2=1$) are located on the dotted black line, with the binary in this configuration at $r = r_{ud+}$ ($r = 10M$) in the bottom-right corner (white circle).
      }\label{F:Lines}
\end{figure*}

The fourth and fifth rows of Fig.~\ref{F:5x5} shows the precession-frequency variation $\Delta\Omega_L$ and the average precession frequency $\langle\Omega_L\rangle$.  Both are correlated with the nutation amplitude $\Delta\theta_L$ shown in the top row because of the features associated with the set of binaries with $\mathbf{J} \parallel \mathbf{L}$ at $r = 10M$.  

\subsection{Role of the $\mathbf{J} \parallel \mathbf{L}$ condition and the up-down configuration}
\label{JparL}

Figure~\ref{F:Lines} shows enlarged versions of the first and fourth rows of the fourth column of Fig.~\ref{F:5x5}.
The dashed lines show binaries for which $\theta_{L-} = 0$ ($\bf{J} \parallel \bf{L}$) as $r/M \to \infty$.  This condition can be expressed analytically by the hyperbola $\chi_1\sin\theta_{1\infty} = q^2\chi_2\sin\theta_{2\infty}$.  The solid lines show binaries for which
$\theta_{L-} = 0$ ($\bf{J} \parallel \bf{L}$) at $r = 10M$.
These lines were determined by setting $\chi_1\sin\theta_1 = q^2\chi_2\sin\theta_2$, $\Delta\Phi_{12} = \pi$ at $r = 10M$, then integrating the precession-averaged radiation reaction backwards in time to determine the asymptotic misalignment angles $\theta_{i\infty}$.  It is fascinating how the gravitational inspiral (in reverse) breaks the symmetry of this analytic condition.

We denote binaries as being in the ``up/down-up/down'' configuration if the primary-secondary spin is aligned (``up'') or anti-aligned (``down'') with the orbital angular momentum $\mathbf{L}$.  The ``up-up,'' ``down-up,'' and ``down-down'' configurations remain stable throughout the inspiral and can therefore be found at the top right, top left, and bottom left corners, respectively, of the $(\cos\theta_{1\infty}-\cos\theta_{2\infty})$ plane.  However, for the parameter choices in Figs.~\ref{F:5x5} and \ref{F:Lines}, the ``up-down'' configuration becomes unstable during the inspiral \cite{2015PhRvL.115n1102G, 2016PhRvD..93l4074L,2020PhRvD.101l4037M,2021PhRvD.103f4003V}.  The binary in the ``up-down'' configuration at $S = S_-$ at $r = 10M$ can instead be found on the conserved dotted line in Fig.~\ref{F:Lines},
\begin{equation} \label{E:chi_eff_line}
(1 + q)\chi_{\rm eff} = \chi_1\cos\theta_{1\infty} + q\chi_2\cos\theta_{2\infty} = \chi_1 - q\chi_2~,
\end{equation}
at the point denoted by the empty circle.
At this point, the ``up-down'' configuration is an unstable equilibrium point on the precession time scale implying $\langle\theta_L\rangle \to 0$ and $\omega \to 0$; this can seen by the white shading at the location of the empty circle in the second and third rows of the fourth column of Fig.~\ref{F:5x5}.  The set of binaries with $\mathbf{J} \parallel \mathbf{L}$ and $\cos\theta_1 > 0$ at $S = S_-$ and $r = 10M$ is marked by the solid black curve connecting the unstable ``up-down'' configuration (empty circle) to the stable ``up-up'' configuration in the top right corner.

The asymmetric inspiral also has the effect of driving binaries with $\mathbf{J} \parallel \mathbf{L}$ and $\cos\theta_1 < 0$ at $S = S_-$ at $r = 10M$ into near complete anti-alignment of the primary spin ($\cos\theta_{1\infty} \simeq -1$) at $r \to \infty$.  This makes the second black curve connecting the ``down-up'' and ``down-down'' configurations nearly indistinguishable from the left edge of the plots (the divergence in $\Delta\Omega_L$ is slightly more pronounced in the third and fifth columns of Fig.~\ref{F:5x5}).

As they inspiral through $r = 10M$, all of the binaries on both of these curves experience:
\begin{enumerate}[label={(\arabic*)}]

\item a local maximum in the nutation amplitude $\Delta\theta_L$\,,

\item a divergence in the precession-frequency variation $\Delta\Omega_L$\,, and

\item a jump in the average precession frequency $\langle\Omega_L\rangle$ by $\pm\omega$.

\end{enumerate}
These features, seen in the first, fourth, and fifth rows of Fig.~\ref{F:5x5}, are the same as those that occur at $r \approx 27M$ for the binary in the right panels of Fig.~\ref{F:Evolutions}.
The numerical results presented in this paper suggest that, when unstable, the ``up-down'' configuration maximizes the nutation amplitude $\Delta\theta_L$ as a function of spin orientation.

\begin{table}
\begin{center} %
\def\arraystretch{1.2}%
\setlength\tabcolsep{0.4em} %
\begin{tabular}{|c|c|c||c|}
 \hline
 $q$ & $\chi_1$ & $\chi_2$ & $f_{\mathbf{J}\parallel\mathbf{L}}$ \\
 \hline\hline
 0.2 & 1 & 1 & 0.02\\
 0.5 & 1 & 0.5 & 0.06\\
 0.5 & 0.5 & 1 & 0.16\\
 0.5 & 1 & 1 & 0.15\\
 0.95 & 1 & 1 & 0.40 \\
 \hline
\end{tabular}
    \caption{The fraction of binaries for which $\mathbf{J}\parallel\mathbf{L}$  at some separation $r>10M$ during the inspiral for BBHs with the same mass ratio $q$ and spin magnitudes $\chi_{1,2}$ as those in Fig.~\ref{F:5x5}.}
\label{T:table1}
\end{center}
\end{table}

BBHs with isotropic spins are uniformly distributed in the ($\cos\theta_{1\infty}-\cos\theta_{2\infty}$) plane.  As shown in Fig.~\ref{F:Lines}, the set of binaries with $\bf{J} \parallel \bf{L}$ is denoted by two curves within this plane that evolve with binary separation from the dashed lines at $r/M \to \infty$ to the solid lines at $r = 10M$.  The fraction of an isotropic population of binaries that pass through such a configuration during the inspiral (and thus experience the three phenomena listed by bullet points in the previous paragraph) is therefore given by the fraction of the area of the ($\cos\theta_{1\infty}-\cos\theta_{2\infty}$) plane bounded by the solid, dashed, and dotted lines in Fig.~\ref{F:Lines}.  This fraction is given in Table~\ref{T:table1} for each of the parameter choices corresponding to the five columns in Fig.~\ref{F:5x5}.  Its increase with mass ratio $q$ can be explained by the following argument.  The binary separation \cite{2015PhRvL.115n1102G,2020PhRvD.101l4037M}
\begin{equation} \label{E:rud+}
\frac{r_{\rm ud+}}{M} = \frac{\left(\sqrt{\chi_1} + \sqrt{q\chi_2}\right)^4}{(1 - q)^2}
\end{equation}
at which the ``up-down'' configuration becomes unstable increases with mass ratio $q$, as does the slope of the line of constant $\chi_{\rm eff}$ given by Eq.~(\ref{E:chi_eff_line}).  This implies that the solid curve in Fig.~\ref{F:Lines} with the white circle (denoting the ``up-down'' configuration at $r = 10M$) as one of its endpoints can migrate further up and to the left, sweeping out more area in the ($\cos\theta_{1\infty}-\cos\theta_{2\infty}$) plane and thus encompassing a higher fraction of binaries.  This is most noticeable in the fourth row, fifth column of Fig.~\ref{F:5x5}, where the endpoint of the curve marking divergences in $\Delta\Omega_L$ has nearly reached the top left corner of the plane.

\subsection{Correlations}
\label{sec:corr}

\begin{figure*}[!t]
  \centering
  \includegraphics[width=\textwidth]{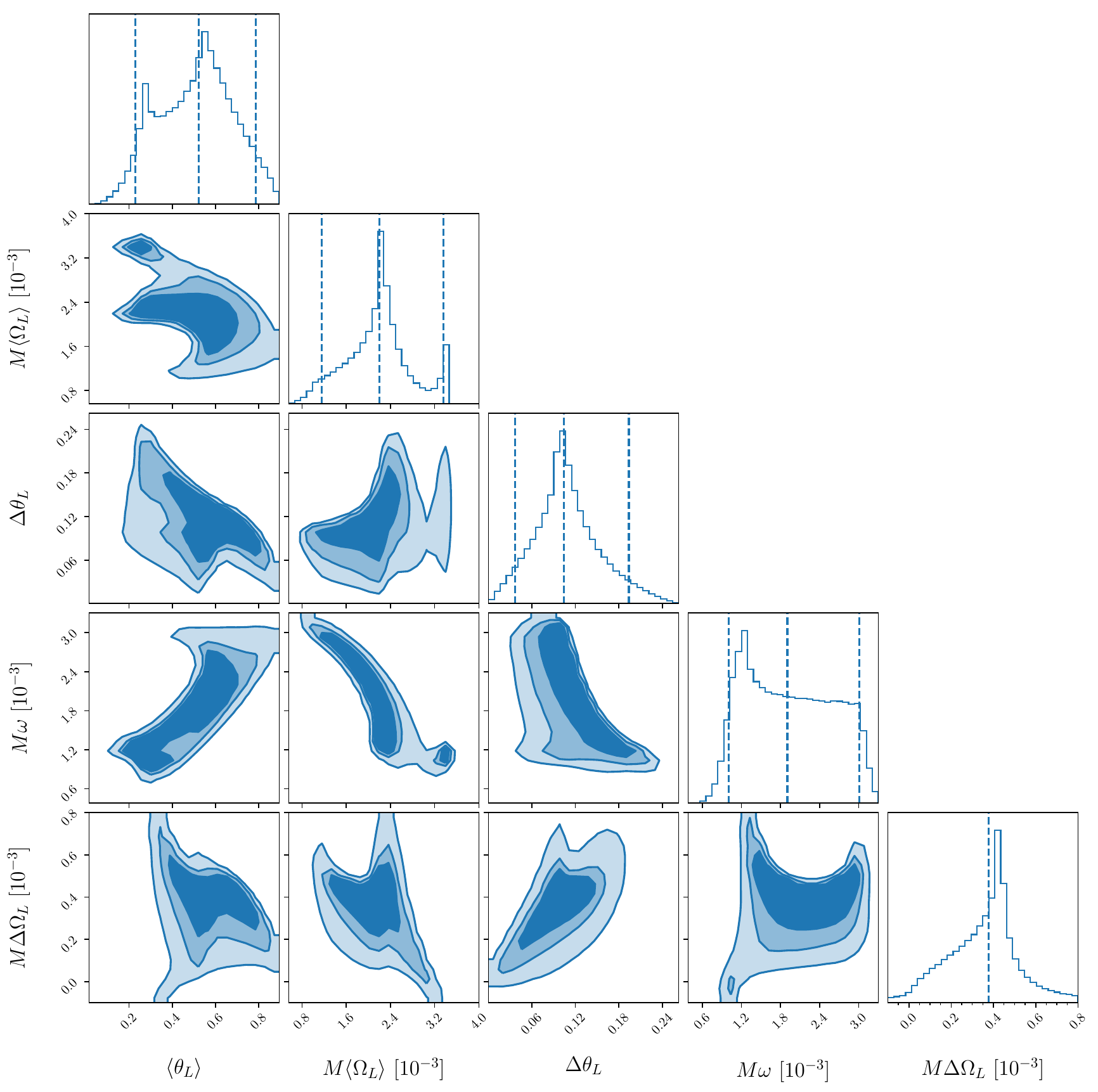}
   \caption{Correlations between the five precession parameters assuming a set of BBHs with $q = 0.5$, $\chi_1 = \chi_2 = 1$, $r = 10M$, and isotropic spins. 2D contour levels encompass $50\%$, $70\%$, and $90\%$ of the BBHs. Medians and $90\%$ intervals of the marginalized distributions are indicated with vertical dashed lines. %
Long tails in the $M\Delta\Omega_L$ distribution have been excluded for clarity. 
}
\label{F:Corner}
\end{figure*}

Fig.~\ref{F:Corner} shows the marginalized 1D and 2D probability distribution functions (PDFs) for our five precession parameters at $r = 10M$ for a population of BBHs with $q = 0.5$, $\chi_1 = \chi_2 = 1$, and isotropic spins.  The 1D PDFs of the average precession amplitude $\langle\theta_L\rangle$ and precession frequency $\langle\Omega_L\rangle$ exhibit distinct bimodality, with subdominant peaks near the $5^{\rm th}$ percentile of $\langle\theta_L\rangle$ and the $95^{\rm th}$ percentile of $\langle\Omega_L\rangle$.  A comparison with the second and fifth rows of the fourth column of Fig.~\ref{F:5x5} reveals that this subpopulation is the $\approx15\%$ of binaries that have passed through alignment of the orbital and total angular momentum ($\mathbf{J}\parallel\mathbf{L}$) during the inspiral.  The jump in $\langle\Omega_L\rangle$ by the nutation frequency $\omega$ as these binaries pass through this alignment is the primary factor that sets this subpopulation apart from the rest of the distribution.  The 2D PDFs of $\langle\Omega_L\rangle$ with the other three precession parameters reveals that this subpopulation disproportionately contributes to the high $\Delta\theta_L$ and low $\omega$ tails (like the unstable ``up-down'' configuration which belongs to the subpopulation).  It dominates the negative $\Delta\Omega_L$ tail, consistent with the behavior seen in the bottom right panel of Fig.~\ref{F:Lines}.

The main BBH population (those $\approx 85\%$ of binaries that never have $\mathbf{J}\parallel\mathbf{L}$ during the inspiral) exhibits many of the correlations previously noted in the discussion of Fig.~\ref{F:5x5}.  There is a positive correlation between $\Delta\theta_L$ and $\Delta\Omega_L$, since both increase as the amount of nutation increases.  Both of these quantities are anti-correlated with the precession amplitude $\langle\theta_L\rangle$, as nutation causes the spins to cancel out in the precession average rather than coherently contribute to misalignment between $\mathbf{J}$ and $\mathbf{L}$.  The nutation frequency $\omega$ is anti-correlated with $\chi_{\rm eff}$ according to Eq.~(\ref{E:dSdt}), implying that it is anti-correlated with $J$ for our isotropic spin distributions.  It is thus correlated with $\langle\theta_L\rangle$ according to Eq.~(\ref{E:ThetaL}) and anti-correlated with $\langle\Omega_L\rangle$ according to Eq.~(\ref{E:Omegaz}).  We also note that the nutation amplitude $\Delta\theta_L$ is anti-correlated with the nutation frequency $\omega$ and correlated with the average precession frequency $\langle\Omega_L\rangle$.  This is primarily driven by the binaries near the solid curve in Fig.~\ref{F:Lines} that have not quite reached $\mathbf{J}\parallel\mathbf{L}$ by $r = 10M$.  Like the unstable ``up-down'' configuration, such binaries have long nutation periods (small $\omega$) during most of which $\Omega_L$ is large because of the smallness of the factor [$(L+S)^2 - J^2$] in the denominator of Eq.~(\ref{E:Omegaz}).

\section{Conclusions} \label{sec:conclusions}

Spin precession is a prominent feature of the relativistic dynamics of BBHs and a key signature of their astrophysical formation channel.
While often simplified using the term ``precession,'' the evolution of the direction of orbital angular momentum $\bf{L}$ is made of a complex superposition of azimuthal (precession) and polar (nutation) motions when defined with respect to a fixed axis such as the direction of the total angular momentum $\bf{J}$. In this work, we have shown that for generic BBHs with misaligned spins, precession and nutation are deeply correlated and occur on the same timescale. Nutation is suppressed only in lower-dimensional regions of the BBH parameter space.

In the construction of gravitational waveforms, the six spin degrees of freedom are often modeled by a reduced set of parameters such as the projected effective spin $\chi_{\rm eff}$~\cite{2008PhRvD..78d4021R,2018PhRvD..98h3007N} and the effective precession spin $\chi_p$ \cite{2015PhRvD..91b4043S,2021PhRvD.103f4067G}.  These parameters aim to capture the dominant spin effects and reduce the computational cost of GW data analysis.  Motivated by the pioneering work of \citeauthor{1994PhRvD..49.6274A}~\cite{1994PhRvD..49.6274A} on the effects of spin precession on gravitational waveforms, we choose a different set of parameters that better characterize the precession and nutation of the orbital angular momentum $\bf{L}$ with respect to the total angular momentum $\bf{J}$.  The five parameters we propose are: the precession amplitude $\langle \theta_L \rangle$, the nutation amplitude $\Delta\theta_L$, the precession frequency $\langle \Omega_L \rangle$, the nutation frequency $\omega$, and the precession-frequency variation $\Delta\Omega_L$. Reference~\cite{2021MNRAS.501.2531S} presented early predictions of the distribution of these parameters in supermassive BBH mergers observable by the LISA mission.

Our numerical investigation indicates that the nutation amplitude $\Delta\theta_L$ is largest for BBHs with:
\begin{enumerate}[label={(\arabic*)}]

 \item moderate mass ratios $q \approx 0.6$,

 \item large spin magnitudes $\chi_i \gtrsim 0.5$, and
 
 \item spin orientations for which $\mathbf{J}\parallel\mathbf{L}$ at some point late in the inspiral.
 
\end{enumerate}
Systems that satisfy condition (3) also experience a divergence in the precession-frequency variation $\Delta\Omega_L$. %
GW events from BBHs satisfying these conditions might offer the best chance to distinguish the effects of precession and nutation and constrain our five parameters observationally.

The next step is to test this hypothesis by exploring the effects of our five precession parameters on the observed gravitational strain $h(t)$.  \citeauthor{1994PhRvD..49.6274A}~\cite{1994PhRvD..49.6274A} investigated how the changing direction of the orbital angular momentum $\bf{L}$ leads to both frequency and amplitude modulation of the gravitational waveform.
Equation (28) of that paper shows that the time derivative of the precessional correction to the orbital phase $\delta\Phi(t)$ is proportional to $d{\hat{\mathbf{L}}}/dt$, which in our notation is given by
\begin{equation}
\frac{d{\hat{\mathbf{L}}}}{dt} = \dot{\theta}_L(\cos\theta_L \hat{\mathbf{L}}_\perp - \sin\theta_L \hat{\mathbf{J}}) + \Omega_L\sin\theta_L(\hat{\mathbf{J}} \times \hat{\mathbf{L}}_\perp)\,.
\end{equation}
We see that nutation ($\dot{\theta}_L \neq 0$) and precession ($\Omega_L \neq 0$) each provide corrections to the orbital and hence GW phase.  They also modulate the GW amplitude 
by introducing time dependence into the factors of $\hat{\mathbf{L}}$ and polarization angle $\psi$ appearing in Eq.~(19a) of \cite{1994PhRvD..49.6274A}.

In a complementary study, \citeauthor{1994PhRvD..49.2658C}~\cite{1994PhRvD..49.2658C} investigated the detectability of the lowest-order spin-dependent correction to the GW phase, shown by \citeauthor{1993PhRvD..47.4183K}~\cite{1993PhRvD..47.4183K} to be proportional to
\begin{eqnarray} \label{E:beta}
\beta &\equiv& \left[ \frac{113}{12}\mathbf{S} + \frac{25}{4}\left(q\mathbf{S}_1 + \frac{1}{q}\mathbf{S}_2 \right) \right] \cdot \frac{\hat{\mathbf{L}}}{M^2} \notag \\
&=& \frac{19}{6} \frac{J\cos\theta_L - L}{M^2} + \frac{25}{4}\chi_{\rm eff}\,.
\end{eqnarray}
Nutation causes $\theta_L$ to oscillate with amplitude $\Delta\theta_L$ and frequency $\omega$, imprinting an additional signature on the GW phase distinct from that of precession.  
Although one ultimately wishes to constrain the magnitudes and misalignments of the individual BBH spins, we hypothesize that the five phenomenological parameters presented in this study can be more tightly constrained because of their more direct connection to the waveform amplitude and phase.
We will explore these signatures of precession and nutation in greater depth %
in an upcoming paper \cite{5Pwave}.

The possibility of measuring our five precession parameters in GW events provides a rich opportunity to identify the astrophysical origin of these systems.  Figure~\ref{F:Corner} shows PDFs of these parameters for an isotropic spin distribution as would be expected for BBHs formed in dynamical interactions in dense clusters.
An upcoming paper \cite{5astro} will explore the distributions of these parameters for BBHs formed from isolated stellar binaries \cite{2021PhRvD.103f3032S}.  As current and future GW observatories discover an increasing number of BBH systems at higher signal-to-noise ratios, the effects of precession and nutation will be detected unambiguously.  We hope that our new precession parameters will aid in the characterization of these systems and help push the frontiers of GW astronomy.

\acknowledgements
We thank Matthew Mould 
for discussions.
D.\,Gangardt and D.\,Gerosa are supported by European Union's H2020 ERC Starting Grant No. 945155--GWmining and Royal Society Grant No. RGS-R2-202004.
D.\,Gerosa is supported by Leverhulme Trust Grant No. RPG-2019-350.
N.S. and M.K. are supported by the National Science Foundation Grant No. PHY-1607031.
E.S. and M.K. are supported by the National Science Foundation Grant No. PHY-2011977.
Computational work was performed on the University of Birmingham BlueBEAR cluster, the Athena cluster at HPC Midlands+ funded by EPSRC Grant No. EP/P020232/1 and the Maryland Advanced Research Computing Center (MARCC).

\appendix

\section{Maximal $\mathbf{L}$ nutations are forbidden} \label{sec:Wide_nutations}

In this Appendix, we prove that the nutation amplitude cannot be maximal, i.e. $\Delta \theta_L<\pi$ for all BBH configurations. Our calculation mirrors that of Ref.~\cite{2019CQGra..36j5003G} for~$\theta_{1,2}$. 

The condition $\Delta\theta_L=\pi$ is possible only if $\cos\theta_{L-} = 1$ and $\cos\theta_{L+} = -1$.
In those cases, Eq.~(\ref{E:ThetaL}) implies
$S_\pm = |J\pm L|$. 
If such a configuration exists, there needs to be values of the constant of motion $J$ and $\chi_{\rm eff}$ which can simultaneously satisfy $S_- = |J- L|$ and $S_+ = J+L$. Using Eq.~(14) of Ref.~\cite{2015PhRvD..92f4016G} (where $\chi_{\rm eff}$ is indicated as $\xi$), these values are 
\begin{equation}
    J^2 = L^2 + \frac{(S_1^2 - S_2^2)(1 - q)}{1+q},
\end{equation}
and 
\begin{equation} \label{eq:app_xi}
    \chi_{\rm eff} = -\left( \frac{r}{M} \right)^{1/2}\,.
\end{equation}
Equation~(\ref{eq:app_xi}) violates the %
limit $|\chi_{\rm eff}|\leq 1$ in the PN regime $r > M$, implying that $\Delta\theta_L$ is strictly smaller than $\pi$ for all physical BBH configurations.

\section{$\Omega_L$ diverges when $\mathbf{L}$ and $\mathbf{J}$ are aligned}
\label{sec: precession frequency diverges}

At 2PN order, the precession vector is given by (e.g.~\cite{2008PhRvD..78d4021R})
\begin{align}
\begin{aligned}
\boldsymbol{\Omega} = \frac{1}{2r^3} \bigg\{ &\left[ 4 + 3q - 3(1+q)\chi_{\rm eff} \left( \frac{M}{r} \right)^{1/2} \right] {\mathbf S}_1 \\
+ &\left[ 4 + \frac{3}{q} - \frac{3}{q} (1+q)\chi_{\rm eff} \left( \frac{M}{r} \right)^{1/2} \right] {\mathbf S}_2 \bigg\}~.
\end{aligned}    
\end{align}
In the limit $\theta_L \to 0$ or $\pi$,
\begin{equation}
{\mathbf S} \cdot \hat{\mathbf{L}}_\perp = 0 \;\;\Longrightarrow \;\;{\mathbf S}_2 \cdot \hat{\mathbf{L}}_\perp = -{\mathbf S}_1 \cdot \hat{\mathbf{L}}_\perp~.
\end{equation}
It follows that
\begin{equation}
\boldsymbol{\Omega} \cdot \hat{\mathbf{L}}_\perp = -\frac{3(1 - q^2) {\mathbf S}_1 \cdot \hat{\mathbf{L}}_\perp}{2qr^3} \left[ 1 - \chi_{\rm eff} \left( \frac{M}{r} \right)^{1/2} \right]~.
\end{equation}
For misaligned spins (${\mathbf S}_1 \cdot \hat{\mathbf{L}}_\perp \neq 0$), this expression does not approach zero as $\theta_L \to 0$ or $\pi$, implying that the second term in Eq.~(\ref{E:OmegaLgeo}) and thus $\Omega_L$ diverges in this limit.

\newpage

\bibliography{bibme}

\end{document}